\documentclass{pasa}

\usepackage{graphicx}
\title[The post-CE X-ray nucleus of NGC~2392]{The post-common-envelope X-ray binary nucleus of the planetary nebula NGC~2392\thanks{Based on observations made with the Mercator Telescope, operated on the island of La Palma by the Flemish Community, at the Spanish Observatorio del Roque de los Muchachos of the Instituto de Astrof\'isica Canarias.}}

\author[Miszalski et al.]{Brent Miszalski$^{1,2,5}$, Rajeev Manick$^{3}$, Hans Van Winckel$^{3}$ and Ana Escorza$^{3,4}$
\affil{$^1$South African Astronomical Observatory, PO Box 9, Observatory, 7935, South Africa}
\affil{$^2$Southern African Large Telescope Foundation, PO Box 9, Observatory, 7935, South Africa}
\affil{$^3$Institute of Astronomy, KU Leuven, Celestijnenlaan 200D, B-3001 Leuven, Belgium}
\affil{$^4$Intitut d'Astronomie et d'Astrophysique, Universit\'e Libre de Bruxelles, ULB, Campus Plaine C.P. 226, Boulevard du Triomphe, 1050 Bruxelles, Belgium}
\affil{$^5$Email: brent@saao.ac.za}
}
\jid{PASA}
\doi{10.1017/pas.\the\year.xxx}
\jyear{\the\year}

\usepackage{aas_macros}

\begin{document}

\begin{frontmatter}
\maketitle

\begin{abstract}
   The \emph{Chandra X-ray Observatory} has detected relatively hard X-ray emission from the central stars of several planetary nebulae (PNe). A subset have no known late-type companions, making it very difficult to isolate which of several competing mechanisms may be producing the X-ray emission. The central star of NGC~2392 is one of the most vexing members, with substantial indirect evidence for a hot white dwarf (WD) companion. Here we report on the results of a radial velocity (RV) monitoring campaign of its central star with the HERMES \'echelle spectrograph of the Flemish 1.2 m Mercator telescope. We discover a single-lined spectroscopic binary with an orbital period of $1.902208\pm0.000013$ d and a RV semi-amplitude of $9.96\pm0.13$ km s$^{-1}$. The high degree of nebula ionisation requires a WD companion ($M\gtrsim0.6 M_\odot$), which the mass-function supports at orbital inclinations $\lesssim$7 deg, in agreement with the nebula orientation of 9 deg. The hard component of the X-ray spectrum may be explained by the companion accreting mass from the wind of the Roche lobe filling primary, while the softer component may be due to colliding winds. A companion with a stronger wind than the primary could produce the latter and would be consistent with models of the observed diffuse X-ray emission detected in the nebula. The diffuse X-rays may also be powered by the jets of up to 180 km s$^{-1}$ and active accretion would imply that they could be the first active jets of a post-common-envelope PN, potentially making NGC~2392 an invaluable laboratory to study jet formation physics. The 1.9 d orbital period rules out a double-degenerate merger leading to a Type Ia supernova and the weak wind of the primary likely also precludes a single-degenerate scenario. We suggest that a hard X-ray spectrum, in the absence of a late-type companion, could be a powerful tool to identify accreting WD companions.
\end{abstract}

\end{frontmatter}

\section{Introduction}
\label{sec:intro}
Identifying the formation mechanisms behind the diverse range of phenomena observed in planetary nebulae (PNe) has proven to be a particularly challenging task (Balick \& Frank 2002). Although binary interactions offer the leading explanation for many phenomena (De Marco 2009; Jones \& Boffin 2017), there are some phenomena that remain particularly enigmatic. The \emph{Chandra} detection of relatively hard X-ray emission with peak energies $>0.5$ keV in some central stars is one such phenomenon (Guerrero et al. 2001, 2011; Guerrero 2012, 2015; Kastner et al. 2012; Freeman et al. 2014; Montez et al. 2015; Montez \& Kastner 2013; Montez 2017). A wide range of explanations have been proposed (see Guerrero 2012 and Montez et al. 2015 for comprehensive discussions) and include processes that can operate from central stars on their own (shocks in the stellar wind, emission from the atmosphere, and accretion of nebular or debris disk material) or in a binary system (colliding winds, accretion onto a companion, and coronal activity from late-type companions). Some detections may be explained by coronal activity only when the presence of late-type stellar companions was already independently established (Montez et al. 2010, 2015). However, in those sources without a proven late-type companion, it is very difficult to determine which of the several possible mechanisms may produce the relatively hard X-ray emission. The most effective means we have to narrow down this list of possible mechanisms is to establish the binary status of those central stars with X-ray detections. 

\begin{figure}
   \begin{center}
      \includegraphics[scale=0.22,bb=0 0 1040.86 1040.86]{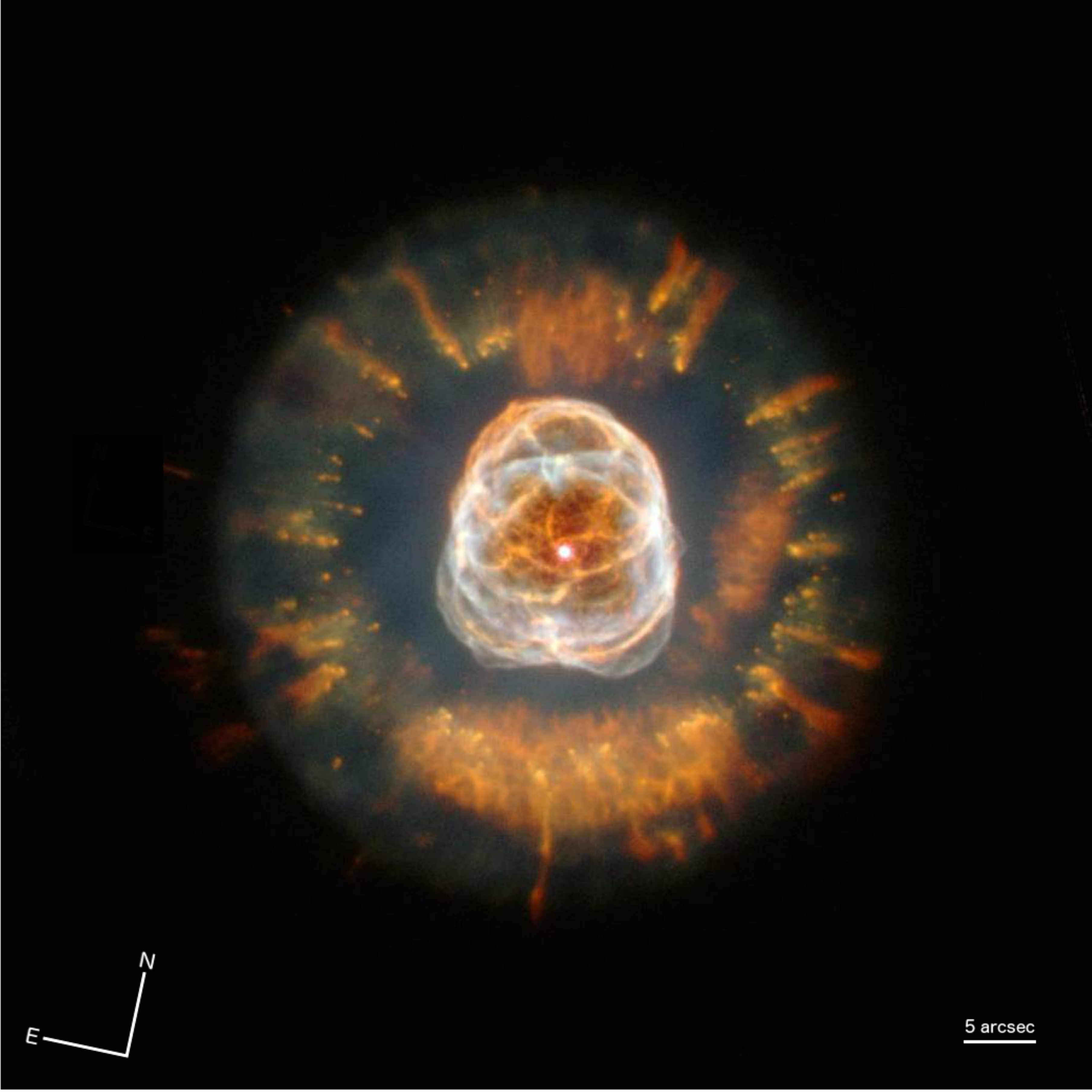}
   \end{center}
   \caption{\emph{Hubble Space Telescope} colour-composite image of NGC~2392 made from images taken with the filters $F469N$ (violet), $F502N$ (blue), $F656N$ (green) and $F658N$ (red). Image credits: NASA, Andrew Fruchter and the ERO Team Sylvia Baggett (STScI), Richard Hook (ST-ECF), Zoltan Levay (STScI).}
   \label{fig:hst}
\end{figure}

The hardest X-ray emission of any central star belongs to NGC~2392\footnote{We refrain from using the common name of NGC~2392 out of respect for indigenous peoples.} (PN G197.8$+$17.3, Fig. \ref{fig:hst}) which exhibits a spectrum peaking above 1 keV and a hard energy tail up to 5 keV (Guerrero 2012; Montez et al. 2015). Guerrero (2012) favoured accretion onto a compact companion to explain the X-ray spectrum. Despite considerable indirect evidence supporting such a companion in the literature, its existence has yet to be proven. Heap (1977) first identified a large discrepancy between the central star effective temperature ($\sim$40 kK) and the He~II Zanstra temperature ($\sim$70 kK), suggesting a hot white dwarf (WD) companion must be present to produce high ionisation potential nebular emission lines like [Ne~V] (see also Pottasch et al. 2008 and Danehkar et al. 2012, 2013). The central star wind has a low terminal velocity of $v_\infty$$\sim$300--400 km s$^{-1}$ (Herald \& Bianchi 2011; Kaschinski et al. 2012; Hoffman et al. 2016), which is much lower than expected to account for the observed diffuse (hot bubble) X-ray emission in the nebula (Steffen et al. 2008; Ruiz et al. 2013). Ruiz et al. (2013) concluded that a companion is required to provide additional energy to power the X-ray emission. Prinja \& Urbaneja (2014) studied short timescale spectroscopic variability in the central star, identifying a possible periodicity of 0.123 d from RV measurements of the N~IV 6380 absorption line, but too few spectra were obtained to determine whether the variability could be due to orbital motion. 

Several nebular properties also favour the presence of a companion. The nebula features a pair of collimated outflows or jets travelling up to 180 km s$^{-1}$ from the central star (O'Dell \& Ball 1985; Gieseking et al. 1985; Garc{\'{\i}}a-D{\'{\i}}az et al. 2012) and numerous low-ionisation structures (LIS, Gon{\c c}alves et al. 2001; Fig. \ref{fig:hst}; Garc{\'{\i}}a-D{\'{\i}}az et al. 2012). Miszalski et al. (2009) first identified jets and LIS to be markers of PNe with post-common-envelope (post-CE) central stars and multiple studies have since strongly corroborated this result with respect to jets (Miszalski et al. 2011a, 2011b; Corradi et al. 2011; Boffin et al. 2012; Tocknell et al. 2014; Miszalski et al. 2018b) and LIS (Corradi et al. 2011; Miszalski et al. 2011b, 2018a; Boffin et al. 2012; Manick et al. 2015; Jones et al. 2014, 2015, 2019; Garc{\'{\i}}a-Segura et al. 2018). Post-CE PNe also seem to exhibit very high abundance discrepancy factors (ADF; Wesson et al. 2018 and ref. therein) and Tsamis et al. (2004) determined an ADF(C$^{2+}$) in the range of 7--24 based on observations by Barker (1991). However, Pottasch et al. (2008) found no discrepancy in the carbon abundance and Zhang et al. (2012) found a low ADF(O$^{2+}$) of 1.65. M\'endez et al. (2012) highlighted that the central star abundances (Pauldrach et al. 2004; Herald \& Bianchi 2011; M\'endez et al. 2012) of He and N are higher than the nebular abundances (see e.g. Pottasch et al. 2008), whereas C and O are lower than in the nebula. M\'endez et al. (2012) interpreted this to be indicative of H-depleted, CNO processed material that was exposed following an interaction with a close binary companion that may have removed the outer layers of the star. 

The indirect evidence supporting a potential binary nucleus in NGC~2392 inspired us to conduct radial velocity (RV) monitoring of its central star. Section \ref{sec:obs} describes the spectroscopic observations and presents the RV measurements. Analysis of the measurements in Section \ref{sec:results} reveals the spectroscopic binary nature of the central star and the orbital parameters are determined. We discuss the results in Section \ref{sec:discussion} in the context of multi-wavelength observations of NGC~2392 and we conclude in Sect. \ref{sec:conclusion}. 

\section{Observations and radial velocity measurements}
\label{sec:obs}
We have obtained 81 observations of the bright central star of NGC~2392 ($V=10.63$ mag, Ciardullo et al. 1999) with the highly-efficient, fibre-fed \'echelle spectrograph HERMES on the Flemish 1.2 m Mercator telescope (Raskin et al. 2011). Table \ref{tab:log} presents a log of the observations which utilised the high-resolution science fibre to give spectra with $R=\lambda/\Delta\lambda=85000$ and a wavelength coverage of 377-900 nm. The data were reduced using the HERMES pipeline (Raskin et al. 2011) which produces spectra with a logarithmic wavelength scale corrected for barycentric motion. The observations were made irregularly with most taken during the period Oct 2015 to Nov 2018, with some also taken during 2010-2011. The Barycentric Julian day in Tab. \ref{tab:log} represents the midpoint of each exposure. The high resolution and stability of the HERMES spectra is ideally suited to detecting low radial velocity amplitudes associated with long orbital period binary central stars of PNe (Van Winckel et al. 2014; Jones et al. 2017). The discovery of spectroscopic binary central stars of PNe via periodogram analysis of multiple RV measurements is now a well established technique (e.g. Boffin et al. 2012; Van Winckel et al. 2014; Manick et al. 2015; Jones et al. 2017; Miszalski et al. 2018a, 2018b). 

The Of(H)-type central star of NGC~2392 (M\'endez 1991) has been extensively studied in the literature (see e.g. Pauldrach et al. 2004; Kaschinski et al. 2012; Herald \& Bianchi 2011; M\'endez et al. 2012; Prinja \& Urbaneja 2014; Hoffmann et al. 2016). Features in the optical wavelength range are mostly due to He and N (see Herald \& Bianchi 2011; Prinja \& Urbaneja 2014) and Prinja \& Urbaneja (2014) found variability on timescales as short as 30 min in many of these features. Here we utilise the N~III $\lambda$4634.14 emission line to measure the RV of the central star in each spectrum. Table \ref{tab:log} records the RV measurements made by fitting models consisting of a Voigt function and a straight line for the continuum to the N~III line profile using the \textsc{lmfit} package (Newville et al. 2016). Figures \ref{fig:fit1}, \ref{fig:fit2} and \ref{fig:fit3} display the resultant fits. The same approach was used by Miszalski et al. (2018b) to succesfully detect orbital motion in the Of(H) central star of the PN MyCn~18. Also included in Tab. \ref{tab:log} are RV measurements of the nebular emission line [O~III] $\lambda$5006.84. The inner core of the complex [O~III] profile was fit with a single Gaussian profile and the mean value of $70.84\pm0.25$ km s$^{-1}$ agrees very well with the nebula systemic velocity of 70.5 km s$^{-1}$ determined by Garc{\'{\i}}a-D{\'{\i}}az et al. (2012).

\begin{table*}
   \centering
   \caption{Log of Mercator HERMES observations of NGC~2392. Radial velocity measurements were obtained from fitting profiles of the emission lines N~III $\lambda$4634.14 (stellar) and [O~III] $\lambda$5006.84 (nebular). }
   \label{tab:log}
   \begin{tabular}{lrlllrll}
      \hline\hline
      Barycentric & Exposure & RV (N~III)    & RV ([O~III]) & Barycentric & Exposure & RV (N~III) & RV ([O~III])    \\
      Julian day    & time (s) & (km s$^{-1})$ &  (km s$^{-1})$ &Julian day& time (s) & (km s$^{-1})$ & (km s$^{-1})$\\ 
      \hline
2455218.54885 & 900 & $69.09\pm0.77$  & $70.73\pm0.34$ & 2457479.46124 & 2700 & $75.95\pm0.94$ & $70.78\pm0.19$    \\
2455231.50234 & 600 & $70.85\pm1.19$  & $70.32\pm0.33$ & 2457488.39625 & 2700 & $76.22\pm0.73$ & $70.94\pm0.18$    \\
2455231.50988 & 600 & $69.65\pm1.32$  & $70.17\pm0.24$ & 2457502.40114 & 2700 & $72.83\pm0.63$ & $70.93\pm0.20$    \\
2455234.52092 & 600 & $88.13\pm2.12$  & $70.44\pm0.20$ & 2457786.57033 & 1755 & $60.60\pm1.09$ & $70.57\pm0.19$    \\
2455298.49559 & 1800 & $61.87\pm0.47$ & $70.97\pm0.15$ & 2457786.59122 & 1755 & $59.97\pm0.94$ & $70.64\pm0.18$    \\
2455502.75288 & 1300 & $83.80\pm0.86$ & $70.83\pm0.22$ & 2457802.61433 & 2700 & $86.07\pm0.65$ & $70.87\pm0.18$    \\
2455508.67155 & 1800 & $75.63\pm1.20$ & $70.72\pm0.23$ & 2457817.51877 & 3500 & $72.14\pm0.65$ & $70.69\pm0.21$    \\
2455545.61769 & 900 & $61.66\pm1.12$  & $71.06\pm0.20$ & 2457838.47995 & 2700 & $72.73\pm0.54$ & $70.98\pm0.18$    \\
2455549.64465 & 900 & $62.48\pm0.98$  & $70.77\pm0.31$ & 2457851.44310 & 2700 & $67.69\pm0.54$ & $71.41\pm0.41$    \\
2455549.65565 & 900 & $62.47\pm1.19$  & $70.95\pm0.26$ & 2457875.41892 & 2489 & $72.74\pm1.37$ & $70.89\pm0.19$    \\
2455549.66665 & 900 & $63.94\pm1.08$  & $71.34\pm0.33$ & 2457893.39033 & 2700 & $69.75\pm0.72$ & $71.04\pm0.23$    \\
2455549.67765 & 900 & $60.48\pm1.27$  & $71.46\pm0.37$ & 2458015.75672 & 2580 & $78.73\pm1.19$ & $71.13\pm0.44$    \\
2455553.62810 & 900 & $69.52\pm0.87$  & $70.96\pm0.29$ & 2458081.67328 & 2700 & $64.48\pm0.54$ & $70.82\pm0.20$    \\
2455553.63910 & 900 & $68.84\pm0.88$  & $70.61\pm0.23$ & 2458088.62666 & 2700 & $65.09\pm0.43$ & $71.26\pm0.26$    \\
2455553.65010 & 900 & $68.69\pm0.94$  & $71.11\pm0.32$ & 2458091.73334 & 2700 & $90.48\pm0.59$ & $70.62\pm0.19$    \\
2455553.66110 & 900 & $69.99\pm0.95$  & $70.51\pm0.23$ & 2458128.61124 & 2700 & $61.18\pm0.39$ & $70.38\pm0.16$    \\
2455553.67210 & 900 & $71.30\pm0.80$  & $70.49\pm0.21$ & 2458166.60739 & 2700 & $60.62\pm0.40$ & $70.50\pm0.32$    \\
2455553.68310 & 900 & $69.98\pm0.82$  & $70.65\pm0.23$ & 2458171.61518 & 2700 & $79.32\pm0.46$ & $70.51\pm0.18$    \\
2455569.65596 & 900 & $75.64\pm0.80$  & $71.33\pm0.17$ & 2458172.46588 & 2700 & $56.81\pm0.57$ & $70.55\pm0.21$    \\
2455569.66736 & 900 & $76.72\pm1.03$  & $70.85\pm0.21$ & 2458182.49052 & 2700 & $76.72\pm0.46$ & $70.65\pm0.19$    \\
2455595.55973 & 1800 & $75.44\pm0.85$ & $70.78\pm0.17$ & 2458185.56409 & 1800 & $62.22\pm0.59$ & $70.59\pm0.19$    \\
2455613.61178 & 1500 & $66.60\pm0.73$ & $71.00\pm0.16$ & 2458185.58551 & 1800 & $59.64\pm0.57$ & $70.75\pm0.18$    \\
2455623.50910 & 1800 & $63.80\pm0.53$ & $71.11\pm0.19$ & 2458188.49635 & 1500 & $72.98\pm0.54$ & $70.64\pm0.17$    \\
2455638.39837 & 2000 & $66.85\pm0.79$ & $70.70\pm0.21$ & 2458188.51429 & 1500 & $72.02\pm0.49$ & $70.68\pm0.16$    \\
2455640.41964 & 2000 & $72.73\pm0.43$ & $70.78\pm0.15$ & 2458190.55543 & 1200 & $78.42\pm0.59$ & $70.49\pm0.21$    \\
2455646.54036 & 900 & $58.96\pm0.93$  & $70.84\pm0.18$ & 2458190.56990 & 1200 & $78.86\pm0.67$ & $70.69\pm0.20$    \\
2457301.66852 & 2700 & $65.20\pm0.54$ & $70.61\pm0.26$ & 2458194.49301 & 2700 & $78.98\pm0.36$ & $70.51\pm0.18$    \\
2457302.69899 & 2700 & $72.44\pm0.56$ & $70.81\pm0.28$ & 2458422.62904 & 1800 & $86.00\pm0.74$ & $71.05\pm0.26$    \\
2457353.58782 & 2700 & $86.22\pm1.46$ & $70.60\pm0.22$ & 2458422.76299 & 1800 & $81.53\pm0.73$ & $71.08\pm0.18$    \\
2457353.65413 & 2700 & $80.66\pm1.41$ & $70.81\pm0.23$ & 2458423.60246 & 1800 & $64.56\pm0.69$ & $70.98\pm0.30$    \\
2457394.66855 & 2700 & $61.99\pm0.72$ & $70.97\pm0.24$ & 2458423.69591 & 1800 & $57.33\pm0.61$ & $71.04\pm0.33$    \\
2457405.67593 & 2700 & $61.40\pm0.76$ & $71.00\pm0.19$ & 2458423.77000 & 1800 & $57.86\pm0.59$ & $71.02\pm0.17$    \\
2457407.67505 & 2700 & $62.30\pm0.72$ & $70.99\pm0.22$ & 2458424.61722 & 1800 & $78.13\pm0.91$ & $70.76\pm0.22$    \\
2457419.57193 & 2900 & $65.69\pm0.92$ & $70.97\pm0.16$ & 2458424.71953 & 1600 & $76.69\pm0.58$ & $70.82\pm0.19$    \\
2457427.52943 & 3100 & $74.63\pm0.73$ & $70.93\pm0.20$ & 2458426.69638 & 1700 & $85.26\pm0.64$ & $70.95\pm0.21$    \\
2457441.44088 & 2700 & $76.02\pm0.89$ & $71.15\pm0.22$ & 2458426.75409 & 1800 & $78.09\pm0.57$ & $70.99\pm0.22$    \\
2457441.47271 & 2700 & $67.43\pm0.79$ & $71.08\pm0.23$ & 2458427.67687 & 1600 & $72.44\pm0.64$ & $70.75\pm0.30$    \\
2457441.56426 & 2700 & $72.31\pm0.77$ & $70.89\pm0.18$ & 2458427.71774 & 1800 & $67.79\pm0.67$ & $71.14\pm0.28$    \\
2457448.59553 & 2700 & $79.84\pm0.86$ & $70.89\pm0.19$ & 2458428.63761 & 1800 & $80.21\pm0.56$ & $71.03\pm0.27$    \\
2457449.38788 & 2700 & $64.59\pm3.80$ & $70.79\pm0.26$ & 2458428.77293 & 1600 & $75.40\pm0.68$ & $70.56\pm0.15$    \\
2457459.49676 & 1060 & $70.13\pm1.85$ & $71.14\pm0.19$ &              &      &                &                    \\
      \hline\hline
   \end{tabular}
\end{table*}

\section{Results and analysis}
\label{sec:results}
\begin{figure}
   \begin{center}
      \includegraphics[scale=0.46,bb=0 0 576 432]{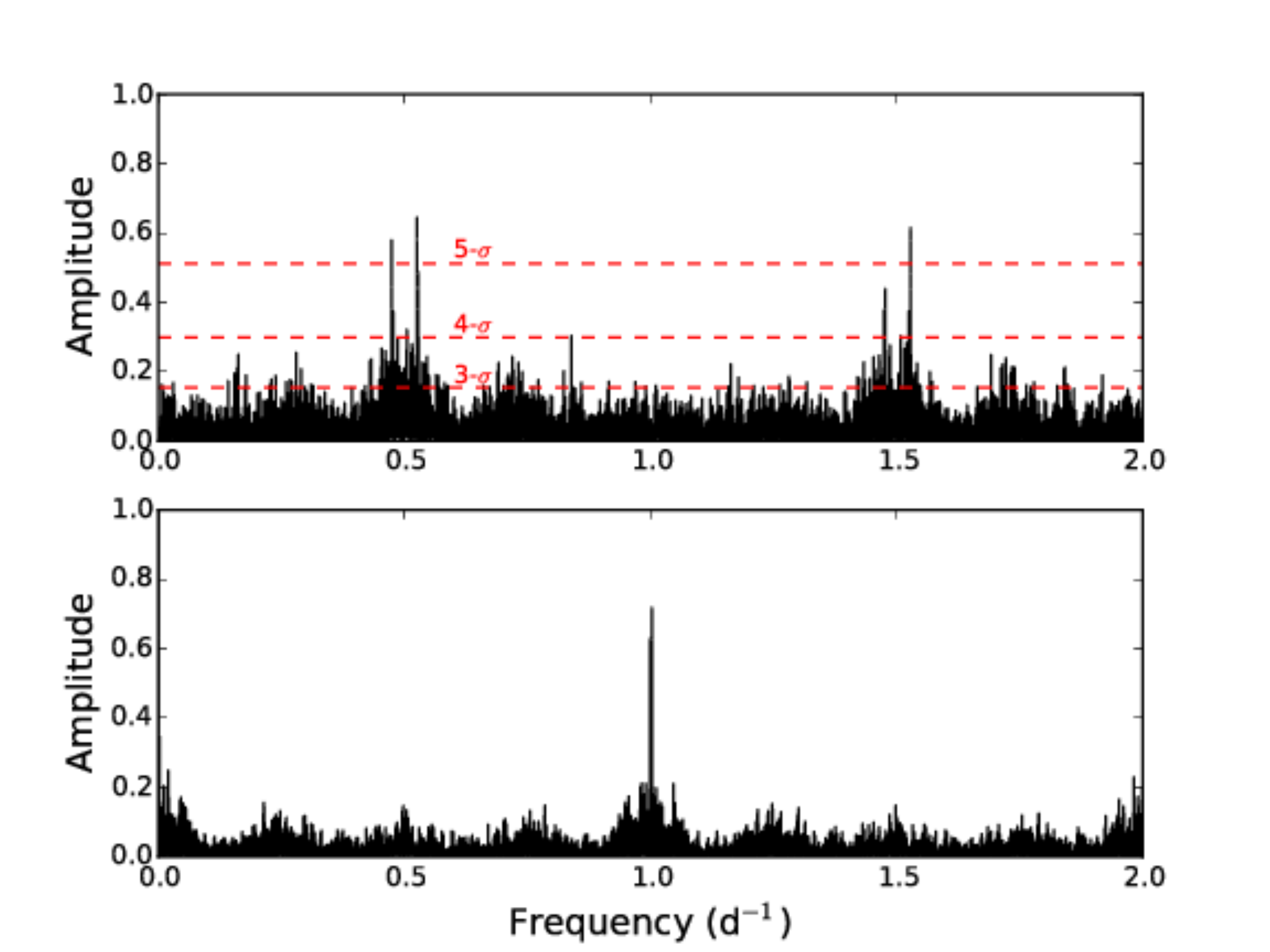}
      \includegraphics[scale=0.35,bb=0 0 724 567]{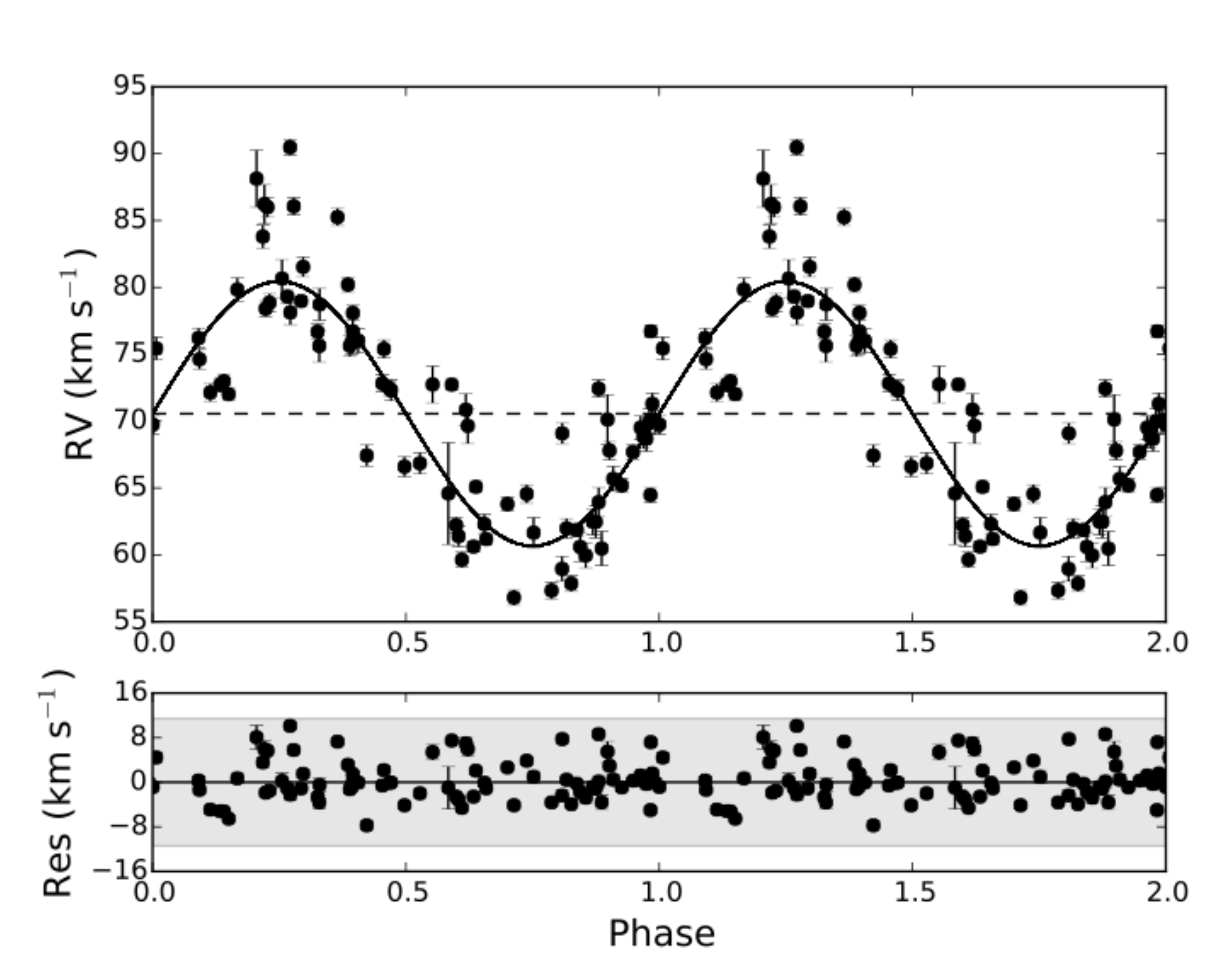}
   \end{center}
   \caption{\emph{(Top panel)} Lomb-scargle periodogram of the HERMES N~III $\lambda$4634.14 RV measurements (top half) and the window function (bottom half). The strongest peak at $f=0.5257$ d$^{-1}$ corresponds to the orbital period and the next strongest peak is its $1+f$ alias. \emph{(Bottom panel)} HERMES RV measurements phased with the orbital period. The solid line represents the Keplerian orbit fit and the dashed line indicates the systemic velocity $\gamma=70.45\pm0.13$ km s$^{-1}$. The shaded region indicates the residuals are within 3$\sigma$ of the fit where $\sigma=3.8$ km s$^{-1}$. }
   \label{fig:results}
\end{figure}

Analysis of the RV measurements in Tab. \ref{tab:log} using a Lomb-Scargle periodogram (Press et al. 1992) reveals the strongest peak to be located at $1.902208\pm0.000013$ d (Figure \ref{fig:results}). This period is significant at greater than the 5$\sigma$ level and the next strongest peak is its $1+f$ alias. We find no other features in the periodogram after subtracting the strongest peak and we do not detect the 0.123 d period reported by Prinja \& Urbaneja (2014).\footnote{The 0.123 d period of Prinja \& Urbaneja (2014) might be due to rotation (e.g. Prinja et al. 2012), but higher cadence observations are required to investigate this further.} We used the 1.9 d period as the basis for a Keplerian orbit model that was built using a least-squares minimisation method applied to the phase-folded data. Figure \ref{fig:results} shows the RV measurements phased with the orbital period, the Keplerian orbit fit and the residuals. 

The data prove the long-suspected binary nature of NGC~2392 and the relatively short orbital period makes it clear that it experienced a common-envelope interaction (Ivanova et al. 2013). Inspection of dynamic spectra phased with the orbital period show all spectroscopic features of the central star follow the motion derived from analysis of N~III RV shifts. No features that could be attributed to a companion are visible. 

Table \ref{tab:orbit} presents the orbital parameters of the Keplerian orbit fit determined using Monte Carlo simulations in which the eccentricity was fixed to be zero (for details, see Miszalski et al. 2018a). The current data do not support an eccentric orbit since the standard Lucy \& Sweeney (1971) diagnostic test results in a probability $p=0.48$, significantly higher than the 0.05 threshold below which an orbit can be considered significantly eccentric. The root-mean-square residuals of the Keplerian fit are relatively high at 3.8 km s$^{-1}$ and this most likely reflects the documented erratic variability of the fast wind on timescales as short as 30 min (Prinja \& Urbaneja 2014). We lack the sufficient temporal coverage to study such variability in greater detail than Prinja \& Urbaneja (2014), but we note that such behaviour is to be expected due to the clumpy and turbulent nature of the fast winds exhibited by central stars of PNe (e.g. Grosdidier et al. 2000, 2001). Comparable levels of residuals are also observed in the binary nuclei of NGC~5189 (Manick et al. 2015; Miszalski et al. 2015) and MyCn~18 (Miszalski et al. 2018b) which also demonstrate fast winds.

The systemic velocity of the orbit ($70.45\pm0.13$ km s$^{-1}$) is in excellent agreement with the the systemic velocity of the nebula (70.5 km s$^{-1}$, Garc{\'{\i}}a-D{\'{\i}}az et al. 2012). The orbital inclination is likely to be close to the nebula orientation of 9 deg determined from spatio-kinematic modelling of the nebula (Garc{\'{\i}}a-D{\'{\i}}az et al. 2012), since other post-CE PNe show remarkably close agreement between the orbital and nebula inclinations (Hillwig et al. 2016). Garc{\'{\i}}a-D{\'{\i}}az et al. (2012) did not provide an estimate of the uncertainty in the nebula orientation. 

\begin{table*}
   \centering
   \caption{Orbital parameters of the best-fitting Keplerian orbit to the N~III $\lambda$4634.14 RV measurements of NGC~2392.}
   \label{tab:orbit}
   \begin{tabular}{lc}
      \hline\hline
      Orbital period (d)        & $1.902208\pm0.000013$\\
      Eccentricity $e$ (fixed)       & 0.00 \\
      Radial velocity semi-amplitude $K$ (km s$^{-1}$)  & $9.96\pm$0.13 \\
      Systemic heliocentric velocity $\gamma$ (km s$^{-1}$) & $70.45\pm$0.13\\
      Epoch at radial velocity minimum $T_0$ (d)     & $2458427.429025\pm$0.000013\\
      Root-mean-square residuals of Keplerian fit (km s$^{-1}$) & 3.8\\
      Separation of primary from centre of mass $a_1\sin i$ (au) & $0.00173\pm$0.00002\\
      Mass function $f(M)$ (M$_\odot$) & $0.00019\pm$0.00001\\
      \hline

   \end{tabular}
\end{table*}

\begin{figure}
   \begin{center}
      \includegraphics[scale=0.55,bb=0 0 415.785625 314.683875]{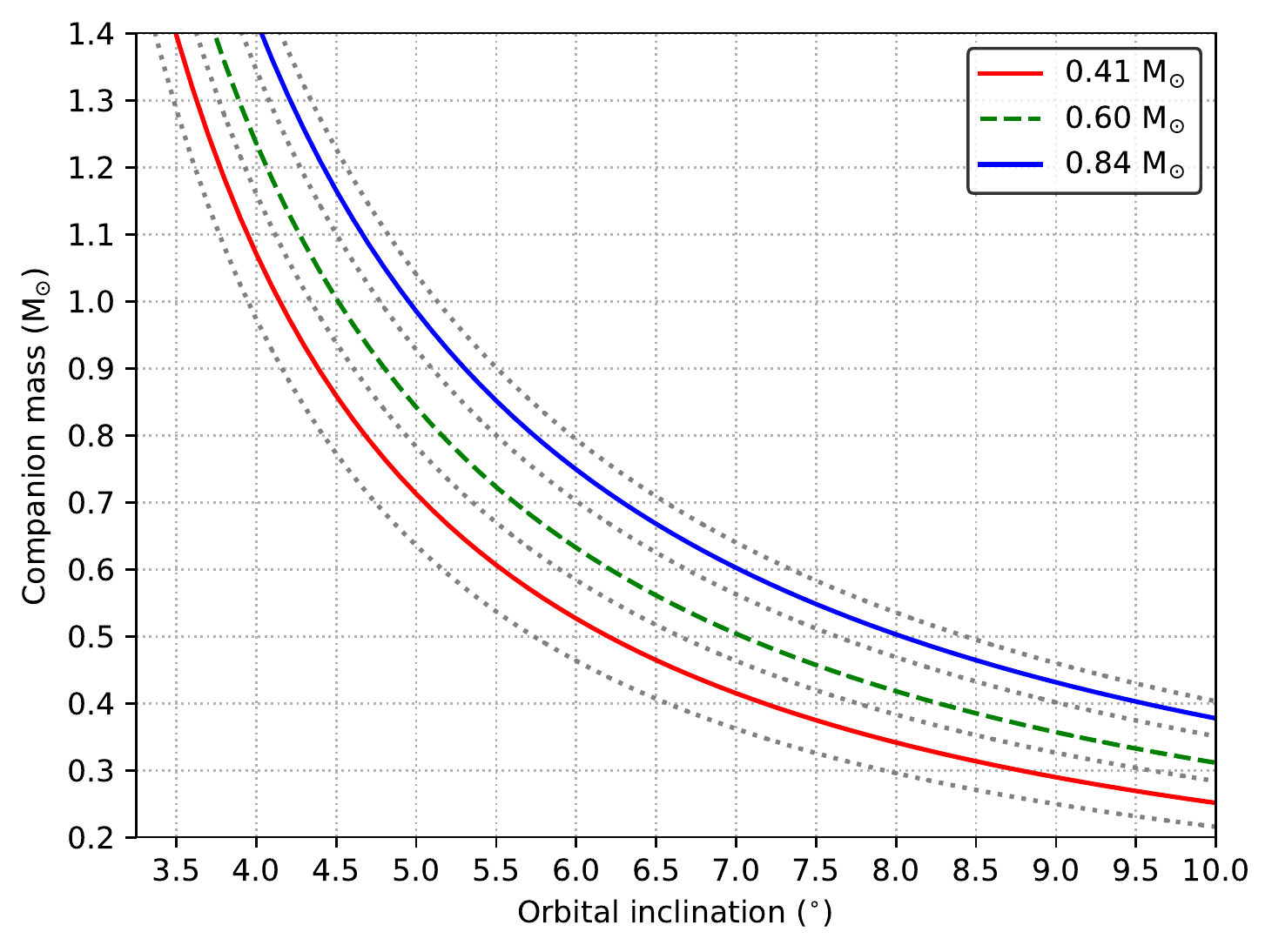}
   \end{center}
   \caption{Companion masses permitted by the mass function in Table \ref{tab:orbit} for primary masses of 0.41 $M_\odot$ and 0.84 $M_\odot$ (Hoffmann et al. 2016), as well as a canonical primary mass of 0.60 $M_\odot$ for reference. The grey dotted lines indicate the companion masses for the 0.1 $M_\odot$ uncertainties in the Hoffmann et al. (2016) masses.}
   \label{fig:masses}
\end{figure}

Figure \ref{fig:masses} depicts companion masses permitted by the mass function for a range of plausible orbital inclinations and primary masses. The primary masses were selected from model atmosphere studies which adopt different strategies to modelling the UV and optical spectra of the central star. The values of $0.41\pm0.10$ and $0.84\pm0.10$ $M_\odot$ appear in Hoffmann et al. (2016), whereas Herald \& Bianchi (2011) assume a canonical 0.6 $M_\odot$. We do not favour any particular value for the primary mass given the complexities associated with the model atmosphere analyses of the central star of NGC~2392 (see Hoffmann et al. 2016 for a detailed review and discussion).

We can also determine whether the primary may experience Roche lobe overflow based on the orbital parameters. This occurs if the radius of the star fills its Roche lobe radius ($R_L$) which Eggleton (1983) defines as:
\begin{equation}
   \frac{R_L}{A} \approx \frac{0.49 q^{2/3}}{0.6 q^{2/3} + \ln(1+q^{1/3})}
   \label{eqn:roche}
\end{equation}
where $A$ is the orbital semimajor axis ($a_1 \sin i$ in Tab. \ref{tab:orbit}) and $q=M_1/M_2$. Adopting radii and masses determined from model atmosphere analyses of the primary, we find that the primary radius $R_1$ does indeed reach $R_L$ for $i\ge4.5$ deg ($M_1=0.41$ $M_\odot$ and $R_1=1.5$ $R_\odot$, Hoffmann et al. 2016), $i\ge4.0$ deg ($M_1=0.84$ $M_\odot$ and $R_1=2.4$ $R_\odot$, Hoffmann et al. 2016) and $i\ge5.0$ deg ($M_1=0.60$ $M_\odot$ and $R_1=1.5$ $R_\odot$, Herald \& Bianchi 2011). It is therefore probable that the primary is currently filling its Roche lobe.

\section{Discussion}
\label{sec:discussion}
\subsection{The WD companion as a possible Type Ia Supernova progenitor}
The presence of highly ionised nebular emission lines such as [Ne~V] (e.g. Pottasch et al. 2008) strictly requires the companion to be a hot WD. Typically this WD would have a mass $\gtrsim$ 0.6 $M_\odot$ and this is satisfied by the mass function for $i \lesssim 7$ deg (Fig. \ref{fig:masses}), with the minimum inclination determined by the Chandrasekhar limit of 1.4 $M_\odot$. It may be possible to further refine the orbital inclination if the low-amplitude photometric variability of Handler (1996, 2003) represents orbital motion, however this may be unlikely given the very low inclination and double-degenerate nature of the binary (e.g. Boffin et al. 2012). 

Alternative means to constrain the companion properties may therefore be required and photoionisation modelling is one approach. Danehkar et al. (2012, 2013) modelled NGC~2392 with a companion mass of $\sim$1.0 $M_\odot$ and effective temperature of 250 kK, leading the authors to suggest that if a close binary nucleus were present, with a total mass exceeding the Chandrasekhar limit, then the double-degenerate (DD) binary could potentially merge to form a Type Ia Supernova (e.g. Maoz et al. 2014). Unfortunately, the 1.9 d orbital period (Tab. \ref{tab:orbit}) firmly rules out this possibility as the binary cannot merge within a Hubble time (Napiwotzki et al. 2005). 

Since the visible star likely fills its Roche lobe (Sect. \ref{sec:results}), there is the possibility that mass transfer can occur onto the WD companion. If the accretion rate of H-rich matter falls within the steady burning regime (e.g. Nomoto et al. 2007), this could potentially lead to a single-degenerate (SD) pathway to a Type Ia supernova (e.g. Maoz et al. 2014). Even though the wind of the visible star is H-rich (Herald \& Bianchi 2011), the mass-loss rate of the visible star does not appear to be high enough to support this outcome (Herald \& Bianchi 2011; Hoffmann et al. 2016), however Wolf-Rayet central stars can have higher mass-loss rates exceeding $\dot{M}\sim10^{-7}$ $M_\odot$ yr$^{-1}$ (Crowther 2008). PNe with central stars that have fast, H-rich winds and WD companions might therefore be promising donors in the SD scenario. Accreting WD companions could be identified via hard X-ray spectra in the absence of late-type companions. 

\subsection{The X-ray spectrum of the binary nucleus}
\label{sec:nucleus}
Previous spectral fitting of the X-ray spectrum of the central star of NGC~2392 involved fitting optically thin thermal plasmas (Guerrero 2012; Montez et al. 2015). The orbital parameters of the newly discovered binary (Sect. \ref{sec:results}) provide some additional insights into the mechanisms behind the X-ray emission. The X-ray spectrum consists of two apparent components: a mostly soft component peaking just after 1 keV and a hard X-ray tail of up to 5 keV (Guerrero 2012). Such a configuration most closely resembles that seen in the X-ray spectra of symbiotic stars (Muerset et al. 1997; Luna et al. 2013).\footnote{Symbiotic stars typically consist of red giants interacting with WDs that occupy a similar part of the Hertzsprung-Russel diagram to central stars of PNe (Miko{\l}ajewska 2004). We may therefore expect there to be some overlap between the X-ray emission properties of symbiotic stars and binary central stars of PNe, especially if the red giant donor wind is substituted by a wind coming from a Of/Of-WR or Wolf-Rayet component, as is the case for NGC~2392. We anticipate this overlap may become clearer as more binary central stars with X-ray emission are discovered and characterised.} In the classification scheme of Muerset et al. (1997), extended by Luna et al. (2013), the soft component of NGC~2392 bears close resemblance to the $\beta$-type sources, whereas the hard component is similar to the $\delta$-type sources. 

The X-ray spectra of $\beta$-types appear soft with energies less than $\sim$2.4 keV and are thought to be caused by the collision of winds from the WD and red giant components (Luna et al. 2013). We do not find unambiguous evidence of colliding winds in our spectra, but colliding winds will be present if models of the diffuse X-ray emission in the nebula are powered by a strong stellar wind from the companion (see Sect. \ref{sec:diffuse}). The high plasma temperature of $3.3\times10^7$ K (Guerrero 2012; Montez et al. 2015) is consistent with the plasma temperatures of $\sim10^7$ K expected for colliding winds (Guerrero 2012). 

In $\delta$-types the hard X-ray component above $\sim$2.4 keV is likely produced in an accretion-disk boundary layer (Luna et al. 2013). Guerrero (2012) favoured accretion onto a compact companion as the main explanation for the X-ray spectrum of the central star of NGC~2392. As the visible star likely fills its Roche lobe (Sect. \ref{sec:results}), the most likely explanation for the hard X-ray tail is indeed accretion, especially when considering the similarity with $\delta$-type X-ray sources in symbiotic stars. If accretion is indeed active in NGC~2392, then we might also anticipate flickering to be present (e.g. Luna et al. 2013). 

\subsection{The diffuse X-ray emission and jets}
\label{sec:diffuse}
\begin{figure}
   \begin{center}
      \includegraphics[scale=0.55,bb=0 0 424.02 316.1448125]{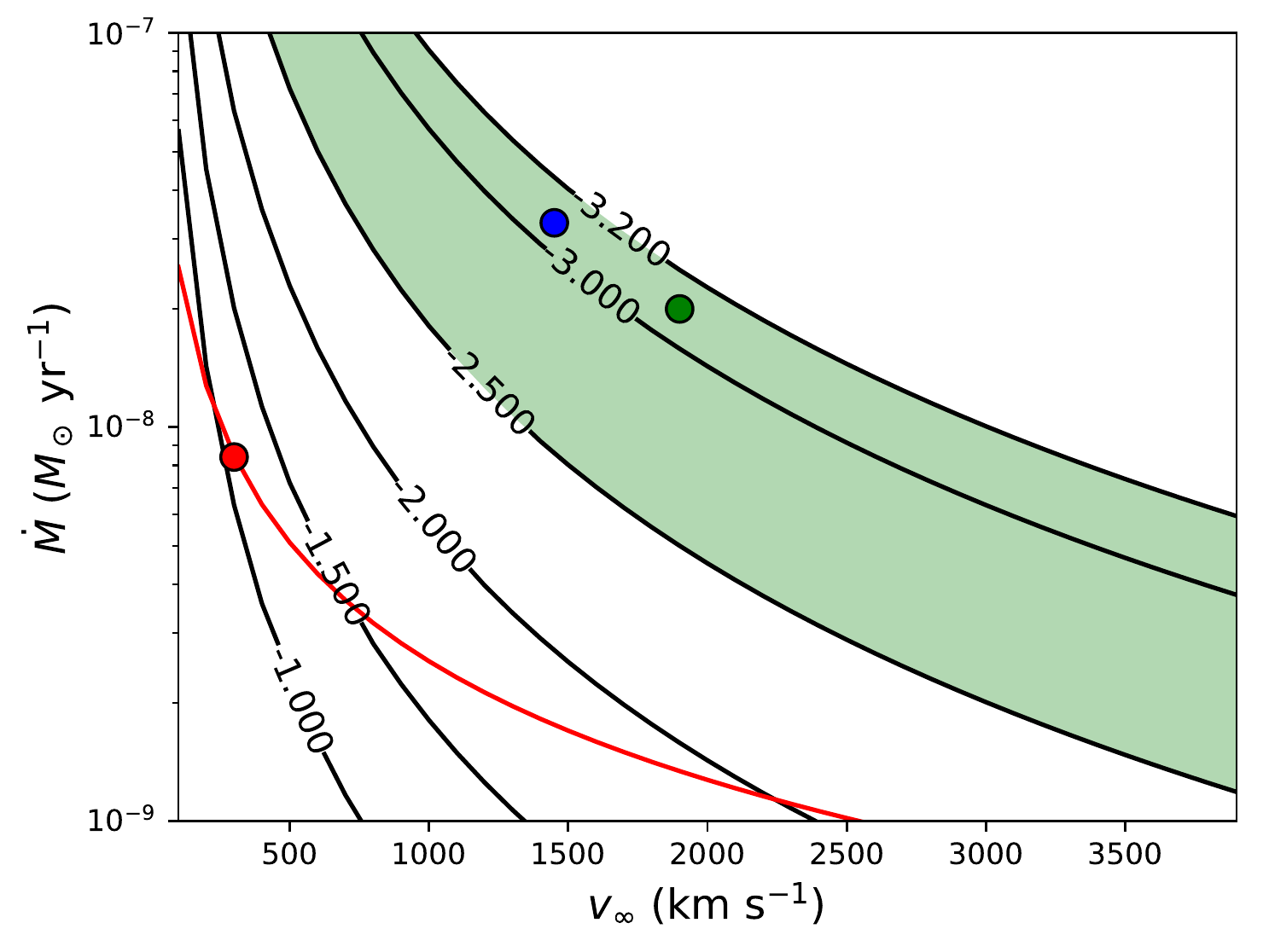}
   \end{center}
   \caption{Possible mass-loss rates ($\dot{M}$) and terminal velocities ($v_\infty$) for the stellar wind of the companion to the central star of NGC~2392. The black contours indicate constant values of $\log L_X/L_\mathrm{wind}$ (Ruiz et al. 2013) and the green shaded region encompass models of Ruiz et al. (2013). If the companion produces the diffuse X-ray emission, its stellar wind is expected to be consistent with the green region. Points indicate the location of the visible central star of NGC~2392 (red, Ruiz et al. 2013) and possible companion locations if we assume it shares the parameters  of the central stars of NGC~6543 (blue, Ruiz et al. 2013) and IC~4663 (green, Miszalski et al. 2012). The red line corresponds to constant $\dot{M}v_\infty$ for NGC~2392.}
   \label{fig:ruiz}
\end{figure}

Spatially resolved, diffuse X-ray emission is observed in the nebula of NGC~2392 (Guerrero et al. 2005; Ruiz et al. 2013) and several other PNe (e.g. Kastner et al. 2012; Freeman et al. 2014). The apparent low wind mechanical luminosity $L_\mathrm{wind}=1/2\dot{M}v_\infty^2$ of NGC~2392 is inconsistent with models of the diffuse emission (Steffen et al. 2008; Ruiz et al. 2013). This is particularly evident when $L_\mathrm{wind}$ is compared against the unabsorbed X-ray luminosity $L_X$ of the diffuse emission in the 0.3--2.0 keV energy range (Fig. 5c of Ruiz et al. 2013). Ruiz et al. (2013) concluded that NGC~2392 was the only object for which the stellar wind of the central star could not explain the diffuse X-ray emission. A binary companion with a higher $L_\mathrm{wind}$ was one explanation proposed by Ruiz et al. (2013). Figure \ref{fig:ruiz} gives an overview of possible values of $\dot{M}$ and $v_\infty$ for the stellar wind of the companion. The contours indicate constant values of $\log L_X/L_\mathrm{wind}$ calculated using the quantities tabulated in Ruiz et al. (2013). 

If the stellar wind of the companion is to power the diffuse X-ray emission, then its position in Fig. \ref{fig:ruiz} must be confined to the green shaded region corresponding to the approximate range covered by the Ruiz et al. (2013) models (i.e. $-3.2 < \log L_X/L_\mathrm{wind} < -2.5$). In this case the companion is likely to have an Of/Of-WR (M\'endez et al. 1990) or Wolf-Rayet (Crowther et al. 1998) spectral type. The companion could therefore potentially resemble NGC~6543 (Herald \& Bianchi 2011) or IC~4663 (Miszalski et al. 2012), both of which have stellar winds that would place them in the acceptable region (Fig. \ref{fig:ruiz}). An expected consequence of such a companion would be the generation of colliding winds, resulting in the formation of a shock cone around the visible central star. This would occur for all companion parameters above the red line in Fig. \ref{fig:ruiz}. The presence of colliding winds would help explain the softer component of the X-ray spectrum of the central star. 

An alternative explanation raised by Ruiz et al. (2013) could be that the jets in NGC~2392 (Garc{\'{\i}}a-D{\'{\i}}az et al. 2012) were responsible for the X-ray emission (Akashi et al. 2008). If we assume that only the jets are responsible, then the requirement for the companion parameters to satisfy the Ruiz et al. (2013) models is relaxed. In this scenario the companion parameters may fall below the red line in Fig. \ref{fig:ruiz} and accretion onto the companion would be easier. A more realistic scenario is one where both the jets and stellar wind contribute to the X-ray emission (e.g. Montez \& Kastner 2018). Asymmetries in the surface brightness distributions of diffuse X-ray emission in PNe (Kastner et al. 2012; Ruiz et al. 2013; Freeman et al. 2014) strongly suggests that jets may help power the diffuse X-rays. 

If accretion is currently active in NGC~2392, as the hard X-ray tail of the central star spectrum suggests, then we could interpret the spatial extent of the jets originating directly from the central star position (Garc{\'{\i}}a-D{\'{\i}}az et al. 2012) as evidence for the first active jets in a post-CE PN. The jets of NGC~2392 could therefore potentially provide direct and unique insights into the mechanisms behind jet formation that other post-CE PNe with dormant jets cannot provide (Tocknell et al. 2014). There is currently no evidence for magnetically confined accretion and spectopolarimetric studies of the visible star find an upper limit of only 450 G for the longitudinal magnetic field (e.g. Steffen et al. 2014). 

In summary, the observed properties of the binary nucleus of NGC~2392 may provide additional valuable constraints on models of diffuse X-ray emission in PNe. The strongest constraints would come from a determination of the stellar parameters of the companion. 

\section{Conclusions}
\label{sec:conclusion}
We have obtained and analysed high-resolution, multi-epoch observations of the central star of NGC~2392 taken with the HERMES spectrograph of the Mercator 1.2 m telescope (Raskin et al. 2011). We summarise our findings as follows:
\begin{itemize}
   \item The Of(H)-type central star of NGC~2392 is a single-lined spectroscopic binary with an orbital period of 1.9 d, proving the long-suspected binary nature of NGC~2392 (e.g. Heap 1977) and making NGC~2392 a post-CE PN. The systemic velocity of the binary ($\gamma=70.45\pm0.13$ km s$^{-1}$) is in excellent agreement with the nebula systemic velocity of 70.5 km s$^{-1}$ (Garc{\'{\i}}a-D{\'{\i}}az et al. 2012). 
   \item Highly ionised nebular emission lines firmly require the companion to be a hot WD (e.g. Pottasch et al. 2008). A typical WD mass of $M_2\gtrsim0.6 M_\odot$ is satisfied by the mass function for $i\lesssim 7$ deg, in good agreement with the nebula orientation of 9 deg determined by spatiokinematic analysis (Garc{\'{\i}}a-D{\'{\i}}az et al. 2012).
   \item Even if the combined mass of both WDs exceeds the Chandrasekhar limit, as suggested by photionisation modelling of the nebula (Danehkar et al. 2012, 2013), the orbital period of 1.9 d excludes the possibility of a binary merger to form a Type Ia supernova (Napiwotzki et al. 2005). 
   \item The radius of the visible central star determined from model atmosphere analyses indicates that it likely fills its Roche lobe radius for $i\gtrsim$4--5 deg. Accretion onto the companion is therefore a plausible origin of the hard X-ray tail up to 5 keV in the X-ray spectrum of the central star (Guerrero 2012; Montez et al. 2015). 
    \item The accretion rate onto the WD companion would likely fall below the steady burning regime required for a SD Type Ia supernova (e.g. Nomoto et al. 2007), however other H-rich central stars with fast winds and WD companions may be promising progenitors for the SD Type Ia supernova channel. In the absence of a late-type companion, a hard X-ray spectrum could be a powerful indicator of an accreting WD companion to other central stars of PNe. 
    \item If the stellar wind of the companion powers the diffuse X-ray emission in the nebula of NGC~2392, then it must have a stronger wind mechanical luminosity than the visible central star. This does not necessarily imply a luminous companion, e.g. IC~4663 (Miszalski et al. 2012) could be hidden against the much brighter visible central star of NGC~2392. This would imply colliding winds are present, which may help explain the softer component of the X-ray spectrum of the central star. 
    \item Alternatively, the jets may contribute most of the power to the diffuse X-ray emission (e.g. Akashi et al. 2008), which is supported by the asymmetric surface brightness distribution of diffuse X-ray emission in many PNe (Kastner et al. 2012; Ruiz et al. 2013; Freeman et al. 2014). In this case the companion may have a weaker wind mechanical luminosity and it would be easier for the companion to host an accretion disk that could launch jets. 
    \item The jets may be the only currently active jets amongst post-CE PNe and their further study could provide unique insights into the jet formation physics of post-CE PNe (e.g. Tocknell et al. 2014).
\end{itemize}

\begin{acknowledgements}
   BM thanks the Institute of Astronomy at KU Leuven for their generous hospitality during the visits in which this work was conducted. We thank Anthony F. J. Moffat for helpful discussions and an anonymous referee for a constructive report. BM acknowledges support from the National Research Foundation (NRF) of South Africa. HVW and RM acknowledge support from the Belgian Science Policy Office under contract BR/143/A2/STARLAB. AE acknowledges support from the Fonds voor Wetenschappelijk Onderzoek Vlaanderen (FWO) under contract ZKD1501-00-W01. HVW acknowledges additional support from the Research Council of K.U. Leuven under contract C14/17/082. IRAF is distributed by the National Optical Astronomy Observatory, which is operated by the Association of Universities for Research in Astronomy (AURA) under a cooperative agreement with the National Science Foundation. 
   
\end{acknowledgements}

\begin{appendix}
   \section{Emission line fits}
   \label{sec:appendix}
\begin{figure*}
   \begin{center}
      \includegraphics[scale=0.79,bb=0 0 634.540625 776.37]{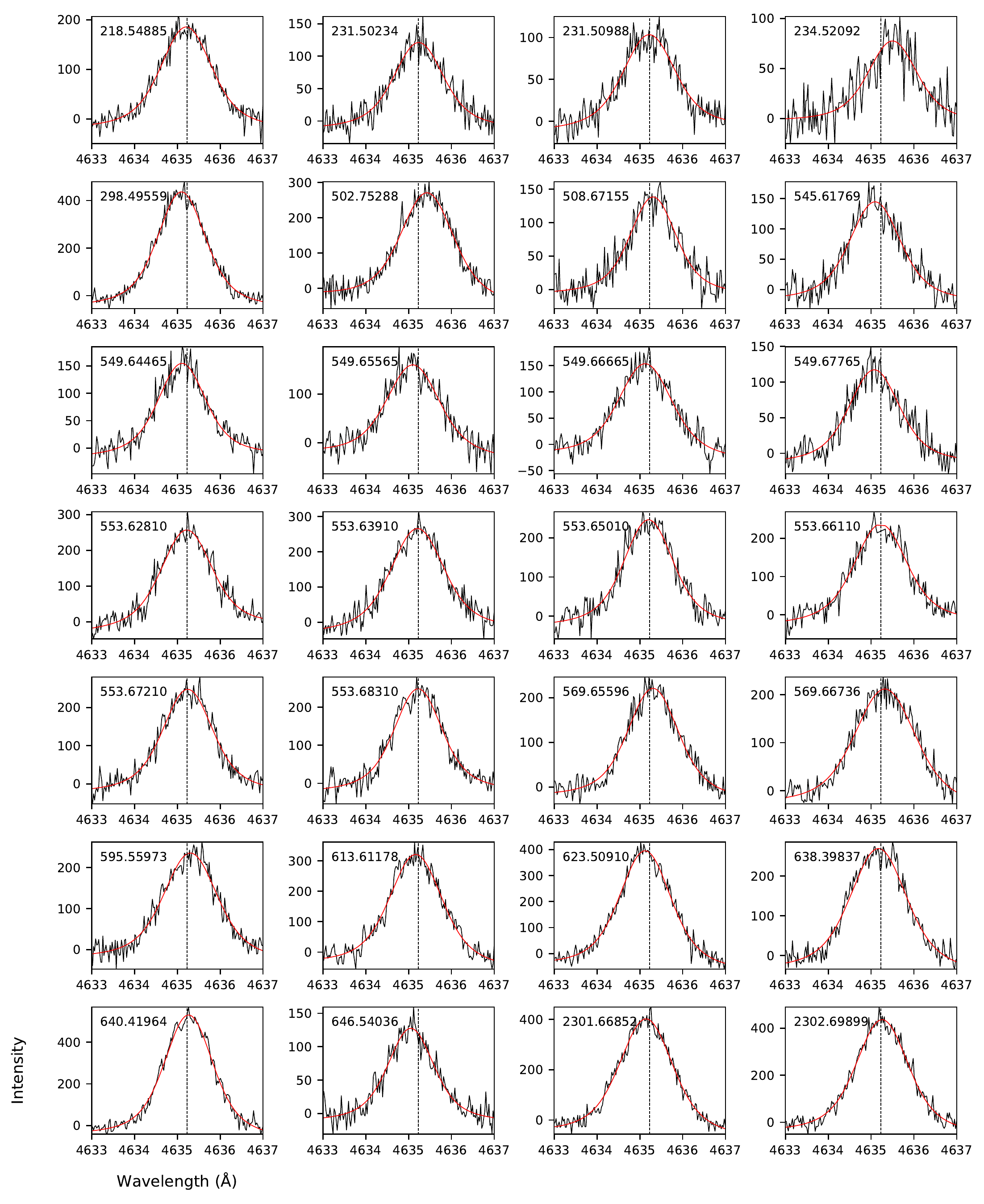}
   \end{center}
   \caption{The observed N~III $\lambda4634.14$ \AA\ profiles (black lines) with the Voigt function and straight line model fits (red lines). The dashed line represents the expected position of N~III at the 70.5 km s$^{-1}$ systemic velocity of the nebula (Garc{\'{\i}}a-D{\'{\i}}az et al. 2012). Each panel is labelled with the Barycentric Julian day minus 2455000 days.}
   \label{fig:fit1}
\end{figure*}

\begin{figure*}
   \begin{center}
      \includegraphics[scale=0.79,bb=0 0 634.540625 776.37]{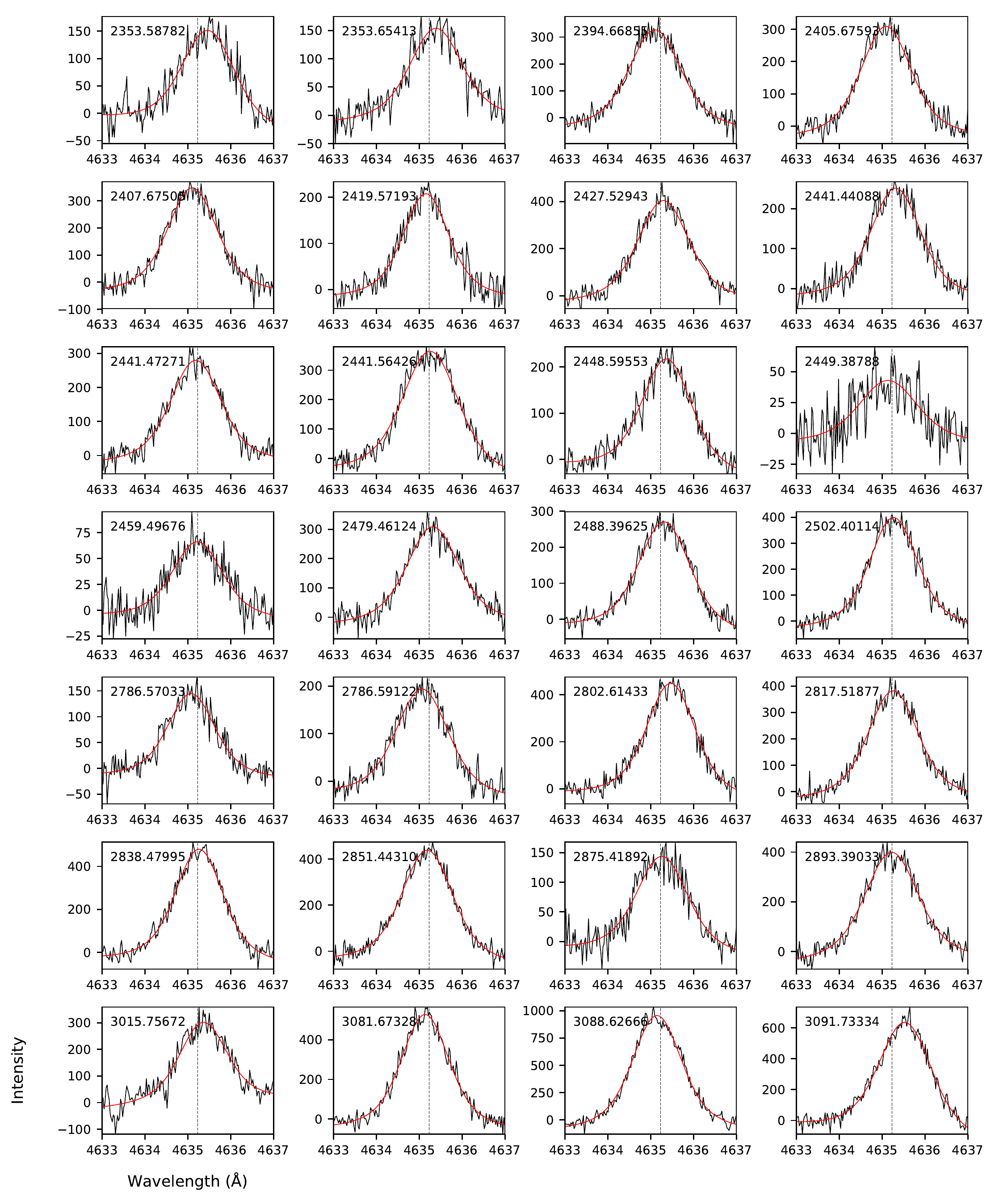}
   \end{center}
   \caption{Figure \ref{fig:fit1} continued.}
   \label{fig:fit2}
\end{figure*}

\begin{figure*}
   \begin{center}
      \includegraphics[scale=0.79,bb=0 0 634.540625 776.37]{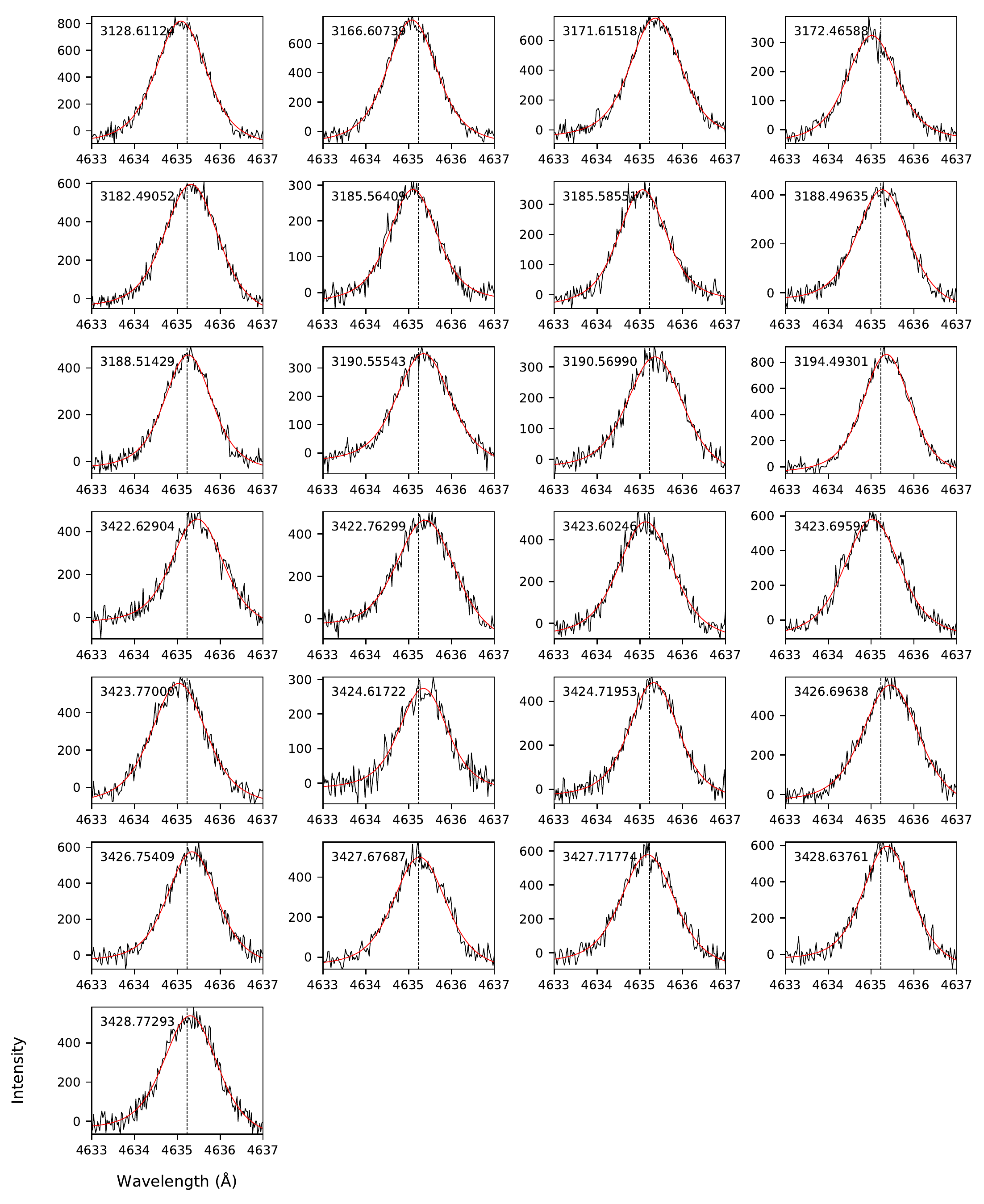}
   \end{center}
   \caption{Figure \ref{fig:fit1} continued.}
   \label{fig:fit3}
\end{figure*}

\end{appendix}

\end{document}